\documentstyle[12pt,epsf]{article}
\newcommand{\nll}{\nonumber \\}

\newcommand{\bq}{\begin{equation}}
\newcommand{\eq}{\end{equation}}
\newcommand{\ba}{\begin{eqnarray}}
\newcommand{\ea}{\end{eqnarray}}

\newcommand{\req}[1]{(\ref{#1})}

\begin{document}
\voffset -2cm
\begin{flushleft}
LMU-09/96
\end{flushleft}
\begin{center}
\vspace{1.5cm}\hfill\\
{\LARGE 
Model-Independent $Z'$ Constraints from Measurements at the $Z$ Peak
}\vspace{1cm}\\
A. Leike\footnote{Supported by the EC-program CHRX-CT940579}\\
Ludwigs--Maximilians-Universit\"at, Sektion Physik, Theresienstr. 37,\\
D-80333 M\"unchen, Germany\\
E-mail: leike@graviton.hep.physik.uni-muenchen.de
\end{center}
%
\begin{abstract}
\noindent
Model independent constraints on extra neutral
gauge bosons are 
obtained from partial decay widths of the $Z_1$ and 
forward--backward and left--right asymmetries at the $Z_1$ peak.
Constraints on the $ZZ'$ mixing angle in $E_6$ models are considered as
special cases. 
 \vspace{1cm}
\end{abstract}
%
%
The measurements at the $Z_1$ resonance at LEP\,1 are in very
good agreement with the SM. 
This agreement can be interpreted as a constraint on new physics.
In particular, the experiments at LEP\,1 give the best present
constraints on the $ZZ'$ mixing angle $\theta_M$
\cite{mixinglim,zmix}. 
Because these constraints depend on the couplings of the $Z'$ to
SM fermions, they are usually given for selected $Z'$ models.
Typical allowed regions for $\theta_M$ in $E_6$ models are
$-0.005<|\theta_M|<0.003$ at the 95\% CL, see \cite{l31961}.
In these {\it model dependent} analyses, all couplings of the $Z'$ to
the SM fermions are linked by model assumptions. 

{\it Model dependent} $Z'$ analyses have the advantage that many observables
can be used as input for the fit.
Furthermore, the resulting $Z'$ constraints can easily be compared
with other experiments. 
They have the disadvantage that the
output is a mixture of experimental data and theoretical assumptions.
For every model, a separate fit has to be performed.

A {\it model independent} $Z'$ analysis does not rely on model assumptions.
It can constrain only certain {\it combinations} of $Z'$ parameters.
Model dependent $Z'$ constraints can be obtained from model
independent $Z'$ constraints as special cases.

For a complete analysis, {\it model dependent} analyses should be
complemented by a {\it model independent} analysis.
Such a model independent $Z'$ analysis is done for off--resonance fermion pair
production \cite{zpmi,lmu0296} and $W$ pair production \cite{pankov} at future
$e^+e^-$ colliders. 
A model independent $Z'$ analysis for fermion pair
production based on present LEP\,2 data also exists \cite{l31961}. 
It is not yet performed for LEP\,1 data.

In this paper, we show how a model independent $Z'$ analysis based on LEP\,1
data could be done using the partial decay widths of the $Z_1$ and
forward--backward and left--right asymmetries at the $Z_1$ peak
as input. 

To fix the notation, we repeat the neutral current interaction with SM
fermions, 
\bq
\label{lagrangian}
{\cal L} = 
e A_\mu J^\mu_\gamma + g Z_\mu J^\mu_Z + g' Z'_\mu J^\mu_{Z'},\ \ \ 
J^\mu_{Z'} = \sum_f \bar f\gamma^\beta(v'_f+\gamma^5 a'_f)f.
\end{equation}

Denoting the symmetry eigenstates by $Z$ and $Z'$, the
mass eigenstates $Z_1$ and $Z_2$ are given by mixing,
\bq
\label{zepmix}
Z_1=Z\cos\theta_M + Z'\sin\theta_M,\ \ \ Z_2=-Z\sin\theta_M + Z'\cos\theta_M.
\eq
In the following, we are interested in the light mass
eigenstate $Z_1$ precisely studied at LEP\,1.
Its axial vector and vector couplings to SM fermions depend on
the $ZZ'$ mixing angle $\theta_M$,
\ba
\label{cmix}
a_f(1)=a_f\cos\theta_M + \frac{g'}{g} a'_f\sin\theta_M,\ \ \ 
v_f(1)=v_f\cos\theta_M + \frac{g'}{g} v'_f\sin\theta_M.
\ea
The eigenvalue equation relates the mixing angle and the masses of the
symmetry and mass eigenstates,
\bq
\label{mmix}
\frac{M_Z^2}{M_1^2} = \rho_{mix} 
= 1+\sin^2\theta_M\left(\frac{M_2^2}{M_1^2}-1\right).
\eq

Consider the partial decay width $\Gamma_1^f$ of the $Z_1$ to $f\bar f$ 
and the forward--backward and left--right asymmetries at the peak in the
limit of small $ZZ'$ mixing,
\ba
\label{peakobs}
\Gamma_1^f &=& M_1\frac{g^2}{12\pi}\left[v_f^2(1)+a_f^2(1)\right]N_f\approx 
\Gamma_1^{f0}\left\{ 1+2\frac{v_fv_f^M+a_fa_f^M}{g(v_f^2+a_f^2)}\right\},\nll
A_{FB}^f &=& \frac{3}{4}A_eA_f
\approx A_{FB}^{f0} + \frac{3}{4}A_f^0\Delta A_e +
\frac{3}{4}A_e^0\Delta A_f,\nll
A_{LR}^f &=& A_f \approx A_f^0 +\Delta A_f
\ea
\bq
\mbox{with\ }A_f\equiv\frac{2a_f(1)v_f(1)}{a_f(1)^2 + v_f(1)^2},\mbox{\ and\ } 
\Delta A_f \approx 2\frac{v_fa_f^M+a_fv_f^M}{g(v_f^2+a_f^2)}-
4\frac{a_f^2v_fa_f^M+v_f^2a_fv_f^M}{g(v_f^2+a_f^2)^2}.
\eq
The index zero denotes the observables without mixing.
We see that measurements at the $Z_1$ peak constrain
the {\it combinations} $a_f^M$ and $v_f^M$,
\bq
\label{coupmdef}
a_f^M\equiv\theta_Mg'a'_f,\ \ \ v_f^M\equiv\theta_Mg'v'_f,
\eq
and not  $a'_f,v'_f$ and $\theta_M$ separately. 
For example, the observables with only leptons in the final state,
$\Gamma_1^l,\ A_{FB}^l$ and $A_{LR}^l$, give model independent constraints on
$a_l^M$ and $v_l^M$. 
Neglecting correlations, they are
\ba
\label{mconstr}
|C_v^lv_l^M+C_a^la_l^M| &<&  \frac{\Delta \Gamma_1^l}{\Gamma_1^l},\nll
|D_v^lv_l^M+D_a^la_l^M| &<& \Delta A_l,\nll
|E_v^lv_l^M+E_a^la_l^M| &<& \Delta A_{FB}^l.
\ea
The coefficients $C_{v,a}^l,D_{v,a}^l$ and $E_{v,a}^l$ depend on SM
parameters only and can be easily deduced from equations \req{peakobs}.
The right hand side of the inequalities \req{mconstr} are experimental errors.

Model independent limits for $a_f^M,v_f^M,\ f=q$ can be obtained
in a similar way with the exception that $A_{FB}^f$ depends on 
the leptonic couplings too. 
We are interested in two--dimensional constraints on $a_f^M$ and $v_f^M$.
Therefore, we interpret the constraint from $A_{FB}^f$ as a second
measurement of $A_f$,
\bq
\label{efferr}
A_f=\frac{4}{3}\frac{A_{FB}^f}{A_e},\ \ \ 
\Delta A_f=\frac{4}{3}\frac{\sqrt{(\Delta
A_{FB}^fA_e^0)^2+(A_{FB}^{f0}\Delta A_e)^2}}{(A_e^0)^2}.  
\eq

To get the connection to real data, radiative corrections have to be
taken into account. They modify the relations \req{mconstr} estimated at
the Born level.
Electroweak corrections can be included
interpreting $a_f$ and $v_f$ in equation \req{cmix} contributing to
$a_f(1)$ and $v_f(1)$ as effective couplings.
Using the effective Weinberg angle $\sin^2\theta_W^{eff}=0.2317$,
excellent agreement with the theoretical predictions in table~2 of
\cite{hollik} is obtained for $\theta_M=0$:
\bq
\label{table}
\begin{array}{lllllll}
{\rm observable}&A_l &A_{FB}^l &A_c &A_{FB}^c &A_b &A_{FB}^b \\
\cite{hollik}   &0.1457   & 0.0159  & 0.667   & 0.0729  & 0.933   & 0.1020 \\
\sin^2\theta_W^{eff}=0.2317&0.1456 &0.0159 &0.667 &0.0728 & 0.935 &0.1022 \\
{\rm experimental\  errors,\ } \cite{warschau}&0.0067 &0.0010 &0.084
&0.0048 &0.049 &0.0023
\end{array}
\eq

We included the electroweak \cite{abr} and QCD corrections for
$\Gamma_1^f$ following \cite{zfitter}, 
\bq
\Gamma_1^f = \frac{N_fG_\mu\sqrt{2}M_1^3}{12\pi}\rho_{mix}\rho_f^Z\mu 
R_{\rm QED}R_{\rm QCD}(M_1^2)\left\{\left[ v_f(1)^2+a_f(1)^2\right]
\left( 1+2\frac{m_f^2}{M_1^2}\right) - 6a_f(1)^2 \frac{m_f^2}{M_1^2}\right\},
\eq
where
\ba
R_{\rm QED} &=& 1 + \frac{3}{4}\frac{\alpha}{\pi}Q_f^2,\ \ \ 
\mu=\sqrt{1-\frac{4m_f^2}{M_1^2}},\nll
R_{\rm QCD} &=& 0\mbox{\ \ for\ }f=l,\nll
R_{\rm QCD} &=& 1 + \frac{\alpha_s(M_1^2)}{\pi} 
+ 1.405\frac{\alpha_s^2(M_1^2)}{\pi^2}
- 12.8\frac{\alpha_s^3(M_1^2)}{\pi^3}
- \frac{Q_f^2}{4}\frac{\alpha\alpha_s(M_a^2)}{\pi^2} \mbox{\ \ for\ }f=q.
\ea
In contrast to \cite{zfitter}, we use the normalization $a_l=-\frac{1}{2}$.
For $\Gamma_1^b$, additional contributions must be taken into account,
see \cite{hollik,zfitter} for original references.

Note that the coupling $g^2$ in equation \req{peakobs} has to be replaced by
$G_\mu$ to meet the experimental accuracy,
\bq
\label{g2repl}
g^2=\frac{4\pi\alpha}{\sin^2\theta_W\cos^2\theta_W}=
\frac{G_\mu\sqrt{2}M_Z^2}{1-\Delta r}
=\frac{G_\mu\sqrt{2}M_W^2}{(1-\Delta r)\cos^2\theta_W},
\eq
where $\Delta r$ is absorbed in $\rho_f^Z$.
The last of the above sequence of equations is valid only for
restricted Higgs sectors.
Through this replacement, a dependence on the $Z$ mass, $M_Z$, appears.
Unfortunately, this induces, through equation \req{mmix}, a dependence
of $\Gamma_1^f$ on $M_2$, which was absent at the Born level. 
For model independent limits on $a_f^M$ and $v_f^M$, this dependence
can be eliminated by additional experimental input.
The knowledge of $M_W,G_\mu$ and $\alpha(M_Z^2)$ together with
equation \req{g2repl} allows to calculate $M_Z$ for models with
restricted Higgs sectors.
The main error of the calculated value of $M_Z$ comes from the
experimental error of the $M_W$ measurement \cite{warschau} and from the  
theoretical uncertainty of $\Delta r$ due to the
unknown Higgs and top mass \cite{hollikzak}. We add these errors quadratically.
As a result, we get $\rho_{mix}=M_Z^2/M_1^2=1 \pm 0.003$. 
We have $\rho_{mix}>1$ in a theory with a $Z'$.
Note that the data used for the calculation of $\rho_{mix}$ are
independent of the measurements used for our further $Z'$ analysis.
We include the {\it shift} of $\rho_{mix}$ from 1, which is within the
experimental errors, in $\rho_f^Z$. 
The {\it error} of $\rho_{mix}$ is included in the error of $\Gamma_1^f$.

\begin{figure}[tbh]
\ \vspace{1cm}\\
\begin{minipage}[t]{7.8cm} {
\begin{center}
\hspace{-1.7cm}
\mbox{
\epsfysize=7.0cm
\epsffile[0 0 500 500]{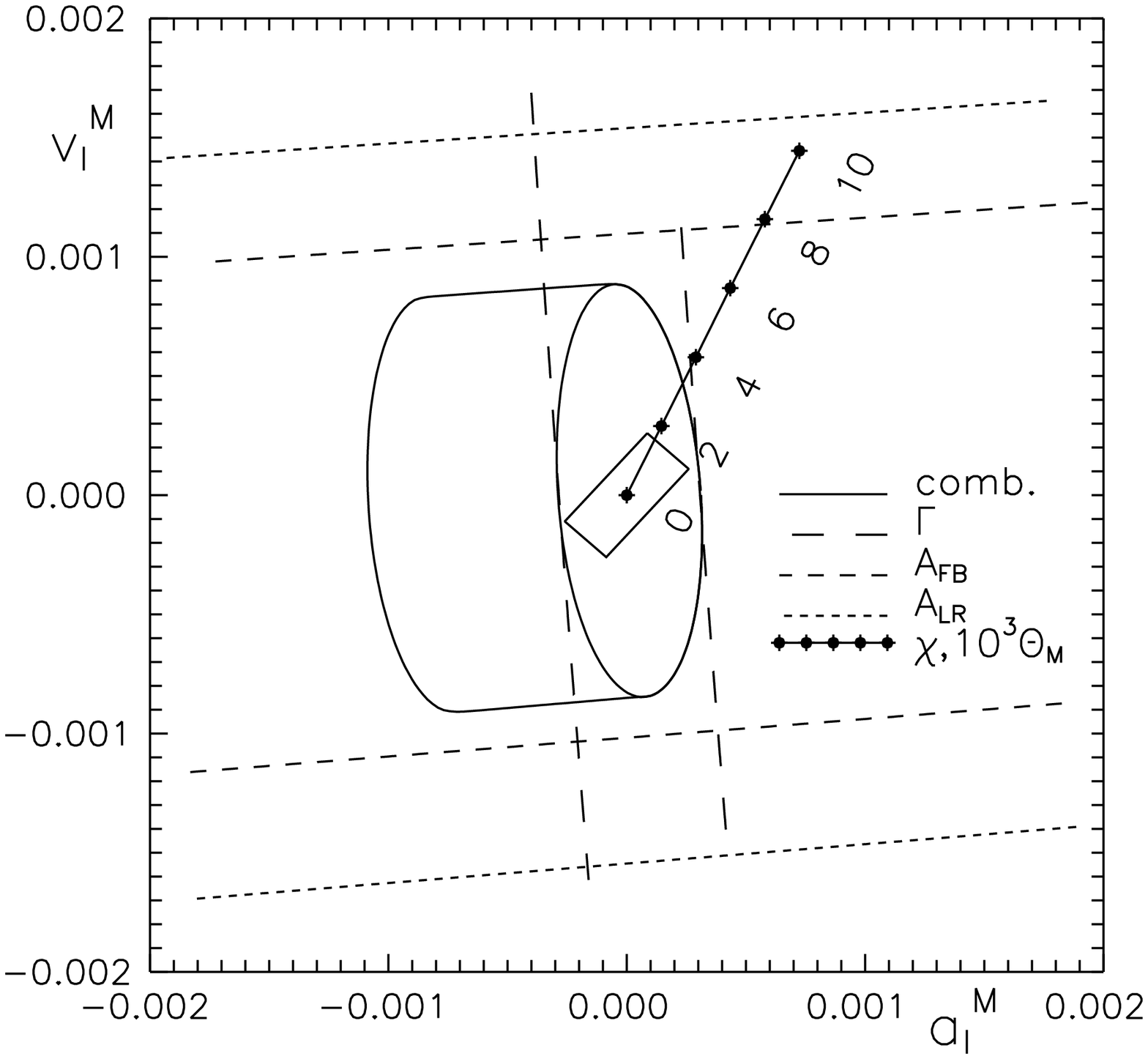}%
}
\end{center}
\vspace*{-0.5cm}
\noindent
{\small\bf Fig.~1: }{\small\it
Areas of $(a_l^M\ v_l^M)$, for which the extended gauge theory's
predictions are indistinguishable from the SM (95\% CL).
Models between the long--dashed (short--dashed, dotted) lines cannot
be detected with $\Gamma_1^l\ (A_{FB}^l,A_{LR}^l)$. 
The regions surrounded by the solid lines cannot be resolved by
all three observables combined, see text.
The numbers at the straight line indicate the value of $\theta_M$ in
units of $10^{-3}$ for the $\chi$ model.
The rectangle is calculated from figure~1 of reference \cite{pankov}.
}
}\end{minipage}
\hspace*{0.5cm}
\begin{minipage}[t]{7.8cm} {
\begin{center}
\hspace{-1.7cm}
\mbox{
\epsfysize=7.0cm
\epsffile[0 0 500 500]{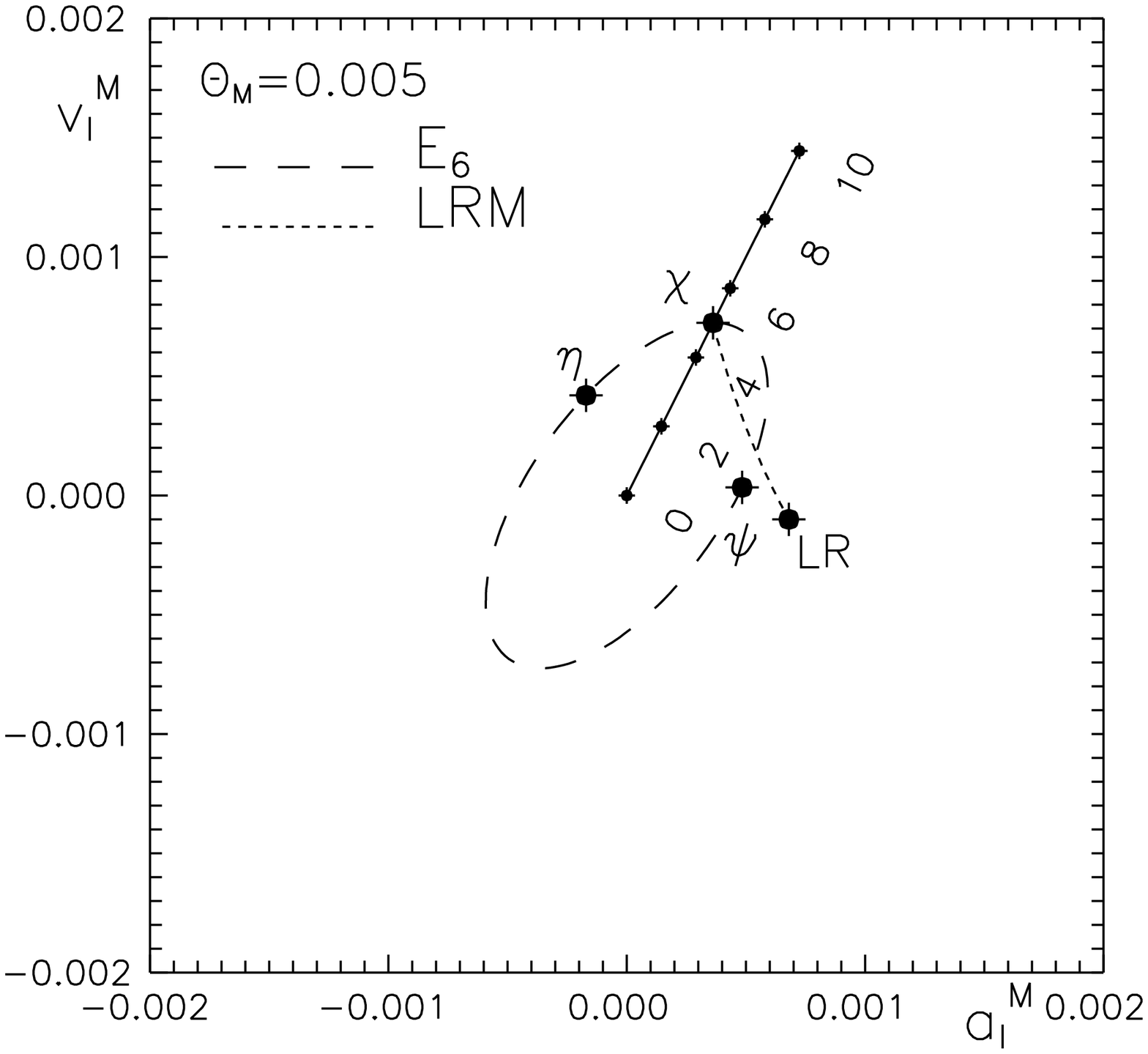}
}
\end{center}
\vspace*{-0.5cm}
\noindent
{\small\bf Fig.~2: }{\small\it 
The vector and axial vector couplings $a_l^M,v_l^M$ for
$\theta_M=0.005$ in typical GUTs. 
For illustration, $\theta_M$ for the $\chi$ model is varied
in units of $10^{-3}$.
}
}\end{minipage}
\end{figure}
%

Figure~1 illustrates the constraints on $a_l^M$ and $v_l^M$.
The plotted regions correspond to $\chi^2=5.99$. 
In our demonstration, the central value of the fit is assumed to be at the
theoretical prediction but the experimental error quoted in
\req{table} is used.
We use $\Gamma_1^l=(83.91\pm 0.11)\,MeV$ \cite{warschau} as input.
Fixing $\rho_{mix}=1$, the region inside the ellipse cannot be
excluded by the data.  
The constraint from every observable is shown separately for $\rho_{mix}=1$.
The uncertainty of $\rho_{mix}$ yields to a shift of the ellipse, which
results to the larger solid region in figure~1.
Future improved measurements of $M_W$ and a determination of $M_H$
would reduce the difference between the two regions.

The {\it model independent} constraints can be interpreted as constraints to
the $ZZ'$ mixing angle $\theta_M$ for any fixed model. 
This is illustrated for the $\chi$ model in figure~1.
Graphically, one obtains $|\theta_M|< 0.0035$. 

Similarly, the constraints on $\theta_M$ could be obtained 
for any other model. 
This is illustrated in figure~2, where $a_l^M,v_l^M$ for $E_6$ models
\cite{e6,e6lr} and left--right models \cite{e6lr,lr} with
$\theta_M=0.005$ are shown. A superposition with figure~1 allows to
obtain the limits on $\theta_M$ for any $E_6$ and left--right model.
The constraints for the $\eta,\psi$ and LR model are
$|\theta_M|<0.01,\ 0.0035$ and 0.0025, correspondingly.
\begin{figure}[tbh]
\ \vspace{1cm}\\
\begin{minipage}[t]{7.8cm} {
\begin{center}
\hspace{-1.7cm}
\mbox{
\epsfysize=7.0cm
\epsffile[0 0 500 500]{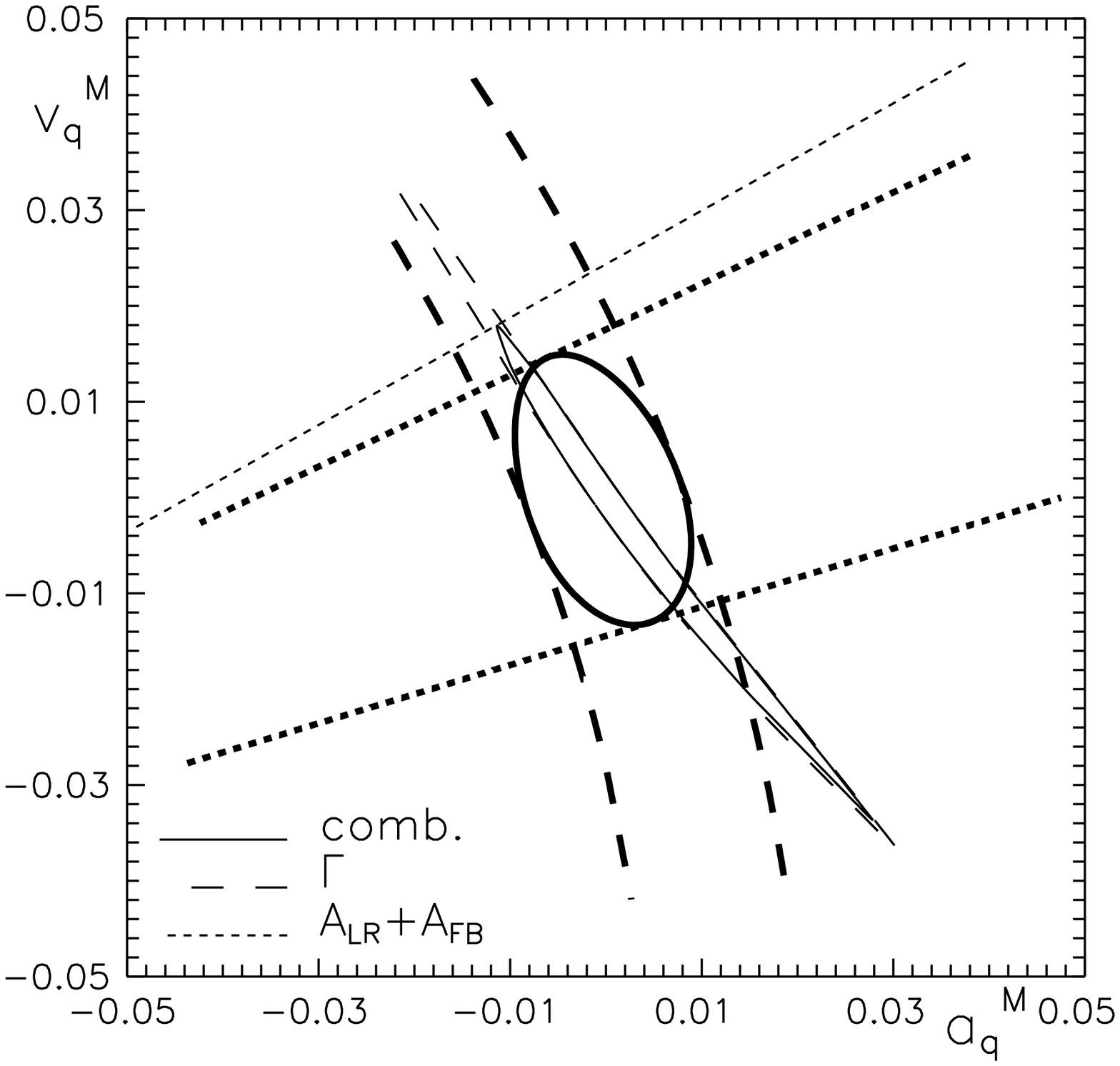}%
}
\end{center}
\vspace*{-0.5cm}
\noindent
{\small\bf Fig.~3: }{\small\it
Areas of $(a_q^M\ v_q^M)$, for which the extended gauge theory's
predictions are indistinguishable from the SM (95\% CL).
Models between the dashed (dotted) lines cannot
be detected with $\Gamma_1^q\ (A_{LR}^q+A_{FB}^q)$. 
The ellipses are the combined regions, which cannot be resolved.
Thick (thin) lines correspond to $q=c(b)$.
}
}\end{minipage}
\hspace*{0.5cm}
\begin{minipage}[t]{7.8cm} {
\begin{center}
\hspace{-1.7cm}
\mbox{
\epsfysize=7.0cm
\epsffile[0 0 500 500]{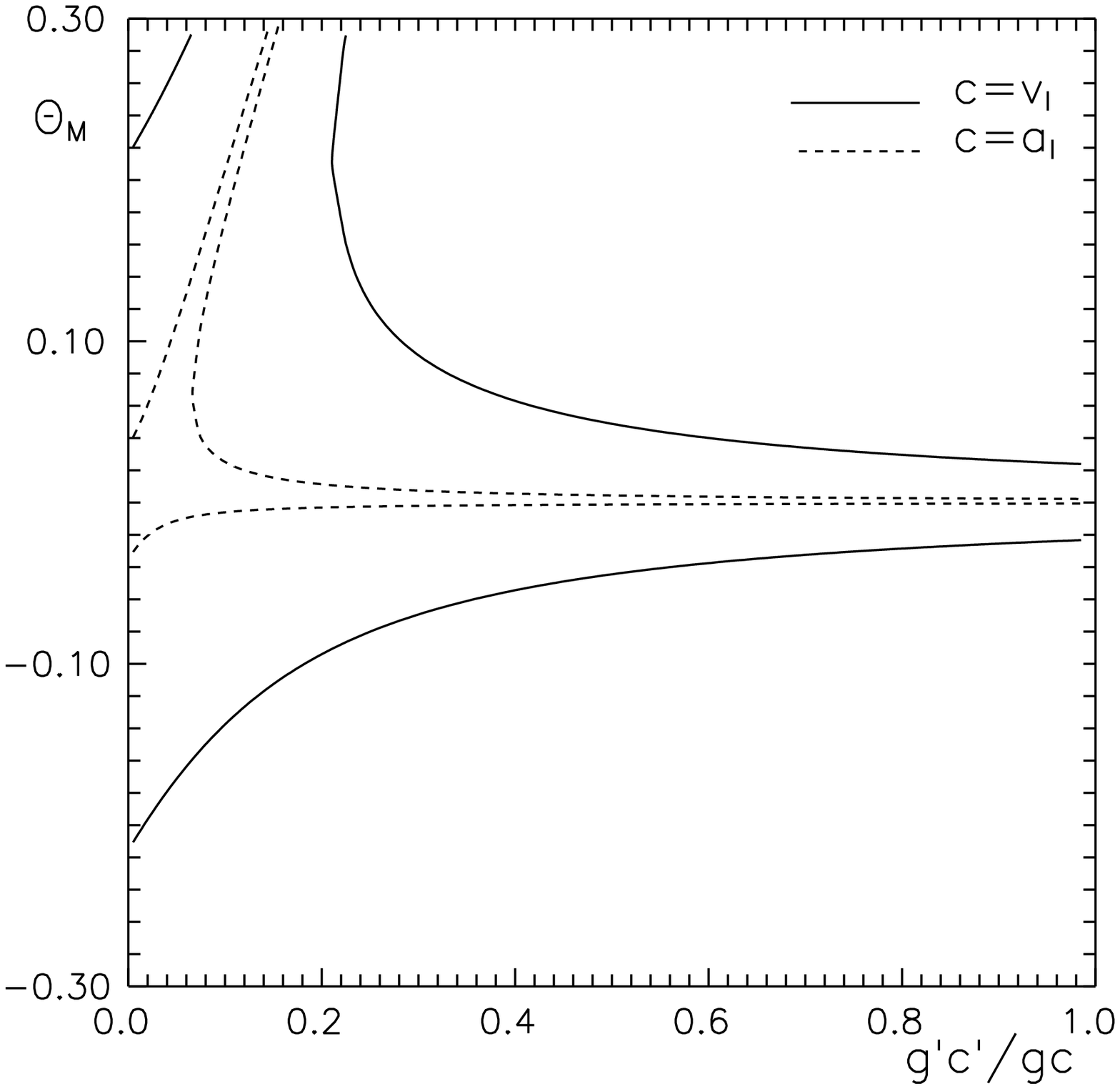}
}
\end{center}
\vspace*{-0.5cm}
\noindent
{\small\bf Fig.~4: }{\small\it 
Areas of $(g'a'_l/(ga_l),\theta_M)$ and $(g'v'_l/(gv_l),\theta_M)$,
for which the extended gauge theory's 
predictions are indistinguishable from the SM (95\% CL).
The region containing the line $\theta_M=0$ cannot be excluded.
}
}\end{minipage}
\end{figure}
%

Figure~3 shows our results similar to figure~1 but for $(a_c^M,v_c^M)$
and  $(a_b^M,v_b^M)$.
The weaker constraints reflect the larger systematic errors of the
quark measurements.
We used $\Gamma_1^b=R_{b}\Gamma_{had}=(379.7\pm1.9)\,MeV$ and
$\Gamma_1^c=R_{c}\Gamma_{had}=(299.0\pm9.8)\,MeV$ as input \cite{warschau}.
In contrast to figure~1, the uncertainty of the exclusion region due
to $\rho_{mix}$ is negligible here.
The missing constraint from $A_{LR}^b+A_{FB}^b$ at one side is due to the
theoretical prediction, which is too close to one to allow for a 
$\chi^2=5.99$ even for $A_b=1$ with the given experimental error.

{\it Model dependent} fits constrain $\theta_M$ using hadronic and
leptonic observables simultaneously. 
Furthermore, there are less $Z'$ parameters to be determined in such a fit.
As a result, the model dependent $Z'$ limits are better than those
obtained for the same model from the model
independent constraints as demonstrated above.
Numerically, the difference can be estimated, for example, comparing
the limit for the $\chi$ model $-0.006<\theta_M<0.008$ taken from
figure~3 of the first reference in \cite{zmix} for $M_{Z'}=700\,GeV$
with the range $|\theta_M|<0.04$, which would be obtained by
our model independent analysis using $\Gamma_1^l$ and $A_{FB}^l$
from the same data set of the analysis \cite{zmix}.

The limits on $a_e^M$ and $v_e^M$ obtained from LEP\,1 data can be
compared with future constraints from $e^+e^-\rightarrow W^+W^-$.
The rectangle in figure~1 is calculated from figure~1 of reference
\cite{pankov} where the constraint on the anomalous couplings $g^*_{WW\gamma}
=1+\delta_\gamma$ and $g^*_{WWZ}=\cot\theta_W+\delta_Z$ is plotted for
$\sqrt{s}=500\,GeV$ and $L_{int}=50fb^{-1}$.  
For small $ZZ'$ mixing and $M_{Z'}^2\gg s$, the values $\delta_\gamma$
and $\delta_Z$ are linearly related to $a_e^M$ and $v_e^M$,
\bq
\delta_Z = \cot\theta_W\cdot\frac{a_e^M}{ag},\ \ \ 
\delta_\gamma=\cot\cdot v_e\theta_W\left(
\frac{a_e^M}{ga}-\frac{v_e^M}{gv}\right)\chi,\ \ \ 
\chi=\frac{s}{s-M_Z^2}.
\eq
The constraint on $a_q^M$ and $v_q^M$ from the $Z_1$ peak cannot be
improved by $W$ pair production. 

The model independent limits in figure~1 can be interpreted as
constraints on a weakly coupling $Z'$.
Explore that figure~1 depends only on the {\it combinations}
$\theta_Mg'a'_l$ and $\theta_Mg'v'_l$. Then, we obtain for
sufficiently small $\theta_M$ the constraint on $\theta_M$ for a $Z'$
model with a ``scaled'' coupling strength $g'\rightarrow \lambda g'$
as $\theta_M\rightarrow \theta_M/\lambda$.
In the simplest approximation where only the linear terms in
$\theta_M$ are kept in equation \req{cmix}, we obtain
\bq
\label{mitheta}
|\theta_M|<\frac{\Delta c}{c}\frac{gc}{g'c'},\mbox{\ \ \ where\ \ \ }
c=a_l,v_l.
\eq
$\Delta c$ is the bound on $a_l^M$ or $v_l^M$ obtained from figure~1,
i.e. $\Delta v_l=0.0008$ and $\Delta a_l=^{+0.0003}_{-0.0011}$.
The exact numerical result is shown in figure~4.
The approximate bound \req{mitheta} is recovered for large $g'c'/(gc)$.
In contrast to \req{mitheta}, the exact calculation gives a bound on
$\theta_M$ also for a $Z'$ with zero coupling, i.e. for $g'c'/(gc)=0$.
It is $|\theta_M|<0.034$ for $c=a_l$.
The existence of such a bound can easily be understood from equation
\req{cmix}, where the deviation of 
the coupling $a_f(1)$ or $v_f(1)$ from $a_f$ or $v_f$ with increasing
$\theta_M$ eventually becomes larger than the 
experimental error, even in the case $g'=0$.
If one allows for a large $ZZ'$ mixing, there is one particular $Z'$
with {\it all} couplings proportional to those of the SM $Z$ boson and
a proper overall coupling strength,
\bq
\label{ssm}
g'c'_f=gc_f\frac{1-\cos\theta_M}{\sin\theta_M},\ c=a,v,
\eq
which would produce no deviation of $c_f(1)$ from $c_f$.
The begin of this second region of insensitivity (besides the region
$\theta_M\approx 0$) can be recognized for small $g'c'/(gc)$ in figure~4.
Such a special model could be detected only by effects of the $Z_2$ propagator.
If equation \req{ssm} is fulfilled for $a'_l$, but not for $v'_l$, the
curve for $v'_l$ in figure~4 would constrain the $ZZ'$  mixing.
Considerations similar to figure~4 can be made for the couplings of
the $Z'$ to quarks. 

To summarize, we demonstrated in a simple analysis how model independent
$Z'$ limits based on the data from the $Z_1$ peak could be obtained.
The constraints on $a_c^M,v_c^M,a_b^M,v_b^M$ are unique, while the
constraints on $a_e^M$ and $v_e^M$ could be improved by measurements
of $e^+e^-\rightarrow W^+W^-$ at energies beyond LEP\,2. 
The relation between model independent and model dependent $Z'$
constraints was discussed.
It was pointed out that LEP data set bounds on the $ZZ'$ mixing angle
even for a $Z'$ with arbitrary small couplings.

In our simple analysis, we neglected correlations between the different
measurements and
assumed that the SM parameters remain unchanged in a $Z'$ fit.
In a more realistic analysis, the SM model parameters and the
model independent $Z'$ parameters $a_f^M,v_f^M$ should be determined
simultaneously and all correlations should be taken into account.
\vspace{0.3cm}

\centerline{\bf Acknowledgments}
I would like to thank W. Hollik, A. Pankov and T. Riemann for discussions.
%


\begin{thebibliography}{99}
\bibitem{mixinglim} 
J. Layssac, F.M. Renard and C. Verzegnassi, Z. Physik {\bf C53} (1992) 97;\\
J. Layssac, F.M. Renard and C. Verzegnassi, Phys. Lett. {\bf B287}
(1992) 267;\\ 
P. Langacker, M. Luo, Phys. Rev. {\bf D45} (1992) 278;\\
L3 collaboration, Phys. Lett. {\bf B306} (1993) 187;\\
G. Altarelli, et al., Phys. Lett. {\bf B318} (1993) 139.
%
\bibitem{zmix}
A. Leike, S. Riemann, T. Riemann, Phys. Lett. {\bf B291} (1992) 187;\\
A. Leike, S. Riemann, Nucl. Phys. {\bf B} (Proc. Suppl.) {\bf 29A}
(1992) 270.
\bibitem{l31961} S. Riemann, L3 Note 1961, 1996.
\bibitem{zpmi} A. Leike, Z. Phys. {\bf C62} (1994) 265.
\bibitem{lmu0296} A. Leike, S. Riemann, preprint DESY 96-111, LMU-02/96, 
hep-ph/9607306, to appear in Z. Phys. {\bf C}.
\bibitem{pankov} A.A. Pankov, N. Paver, preprint IC/96/175,
UTS-DFT-96-01, hep-ph/9610509.
\bibitem{hollik}
W. de Boer, A. Dabelstein, W. Hollik, W. M\"osle, U. Schwickerath,
IEKP-KA/96-08 (revised version), hep-ph/9609209.
\bibitem{warschau} A. Blondel, Plenary talk at 28th Int. Conf. on High
Energy Physics, Warsaw, July 1996; LEP Electroweak Working Group,
CERN-preprint LEPEWWG/96-02.
\bibitem{abr} A. Akhundov, D. Bardin, T. Riemann, Nucl. Phys. {\bf
B276} (1986) 1.
\bibitem{zfitter} C. Bardin, et al., ZFITTER description, CERN-TH. 6443/92.
\bibitem{hollikzak} W. Hollik, Karlsruhe preprint KA-TP-24-1996.
\bibitem{e6} For a review see e.g.
             J.L. Hewett, T.G. Rizzo, Phys. Rep. {\bf 183} (1989) 193.
\bibitem{e6lr} L.S. Durkin, P. Langacker, Phys. Lett. {\bf B166} (1986) 436.
\bibitem{lr} For a review see e.g.
             R.N. Mohapatra, ``Unifications and Supersymmetries'', (Springer,
             New York, 1989).
\end{thebibliography}
\end{document}